\documentclass[reprint,superscriptaddress,amsmath,amsfonts,amssymb,aps,prl,showkeys]{revtex4-1}
\usepackage{bm}
\usepackage{color}
\usepackage{graphicx}
\usepackage[normalem]{ulem}
\usepackage[makeroom]{cancel}
\usepackage{lipsum}

\newcommand{\x}[1]{\text{#1}}

\usepackage{hyperref}
\hypersetup{
  breaklinks = {true},
  citecolor = {blue},
  colorlinks = {true},
  linkcolor = {red},
  urlcolor = {blue},
}

\graphicspath{{../Figs/}}

\begin{document}


\title{Giant Spin Transport Anisotropy in Magnetic Topological Insulators}

\author{Marc Vila}
\affiliation{Department of Physics, University of California, Berkeley, California 94720, USA}
\affiliation{Materials Sciences Division, Lawrence Berkeley National Laboratory, Berkeley, California 94720, USA}

\author{Aron W. Cummings}
\affiliation{Catalan Institute of Nanoscience and Nanotechnology (ICN2), CSIC and BIST, Campus UAB, Bellaterra, 08193 Barcelona, Spain}



\author{Stephan Roche}
\affiliation{Catalan Institute of Nanoscience and Nanotechnology (ICN2), CSIC and BIST, Campus UAB, Bellaterra, 08193 Barcelona, Spain}
\affiliation{ICREA--Instituci\'o Catalana de Recerca i Estudis Avan\c{c}ats, 08010 Barcelona, Spain}

\date{\today}

\begin{abstract}

We report on exceptionally long spin transport and giant spin lifetime anisotropy in the gapped surface states of three-dimensional (3D) magnetic topological insulators (MTIs). We examine the properties of these states using the Fu-Kane-Mele Hamiltonian in presence of a magnetic exchange field. The corresponding spin textures of surface states, which are well reproduced by an effective two-band model, hint at a considerable enhancement of the lifetime of out-of-plane spins compared to in-plane spins. This is confirmed by large-scale spin transport simulations for 3D MTIs with disorder. The energy dependence of the spin lifetime anisotropy arises directly from the nontrivial spin texture of the surface states, and is correlated with the onset of the quantum anomalous Hall phase. Our findings suggest novel spin filtering capabilities of the gapped surface MTI states, which could be explored by Hanle spin precession measurements.
\end{abstract}

\maketitle

%

Topological insulators (TIs) and magnetic topological insulators (MTIs) are two classes of materials with nontrivial topological order that have attracted enormous interest for their rich physics \cite{RevModPhys.82.3045, RevModPhys.83.1057, Tokura2019, Bernevig2022, Liu2023}. The breaking of time reversal symmetry in MTIs due to the presence of magnetic atoms or proximity-induced magnetism yields the long sought-after Chern insulating phase, exhibiting the quantum Anomalous Hall effect (QAHE), predicted by Duncan Haldane \cite{PhysRevLett.61.2015}. MTIs provide an ideal arena for the exploration of topologically distinct phenomena and could be used in magnetoresistive applications for advancing dissipationless topological electronics \cite{Yu2010, Chang2013,Tokura2019, Götte2016ER, An2021, Paul2021}.

A major difference between TIs and MTIs lies in the peculiar properties of their topological states, which naturally reflect the presence or absence of time reversal symmetry. While spins in TI surface states are known to be locked to the momentum, leading to many relevant phenomena in spintronics \cite{Garate2010, Yazyev2010, Mellnik2014, Kondou2016, Khang2018, Farzaneh2020, He2021}, the spin degree of freedom in the topological states of MTIs has received less attention \cite{Wu2014, Zhang2016}. Only recently have spin transport experiments been suggested to reveal signatures of the MTI chiral edge states \cite{PhysRevLett.126.167701}. In fact, to date, there has been no in-depth exploration of spin transport in bulk MTI materials or their topological states. There are for instance no experimental reports of Hanle spin precession or nonlocal spin valves realized in the otherwise well-characterized family of MTIs such as ${\rm MnBi_{2}Te_{4}}$ \cite{Otrokov2019, PhysRevLett.122.107202}. Such measurements allow to probe spin relaxation times and lengths and can be used to extract useful information about the underlying spin texture \cite{Cummings2017, Benitez2018, Ghiasi2018, Garcia2018}, the strength of spin-orbit fields \cite{Vila2021} or the presence of induced magnetism \cite{Karpiak2020, Ghiasi2021}.

One possible reason behind the lack of spin transport studies in MTIs could be the belief that magnetization or strong spin-orbit coupling will likely relax the spins very quickly, resulting in small spin lifetimes and signals. Nevertheless, understanding the applicability and signatures of these experimental techniques in MTIs will be fundamental for the exploration and characterization of the quickly-growing number of new MTIs being grown and discovered \cite{Tian2020, Xu2020, Frey2020, Kamal2021}. First, such understanding may provide alternative experimental approaches to identifying MTI phases in complex three-dimensional materials or heterostructures, and second it may eventually harness such topological materials for novel spin device functionalities.

In this Letter, we use large-scale numerical simulations to study spin transport in three-dimensional models of magnetic topological insulators, including the effect of disorder. We find that the gapped surface states exhibit a large spin lifetime anisotropy between in-plane and out-of-plane spins that is strongly energy dependent, which arises as a consequence of the interplay between the spin texture of the TI surface states modified by magnetic exchange. An important consequence of such interplay is that the out-of-plane spins do not fully relax, suggesting a possible experimental fingerprint of MTIs in spin transport experiments. This behavior serves not only as a resource to identify the strength of magnetism competing with intrinsic spin-orbit coupling in MTIs, but could also serve as a means to design novel spin-dependent functionality such as spin filters.

\begin{figure}[t] 
\centering
\includegraphics[width=0.48\textwidth]{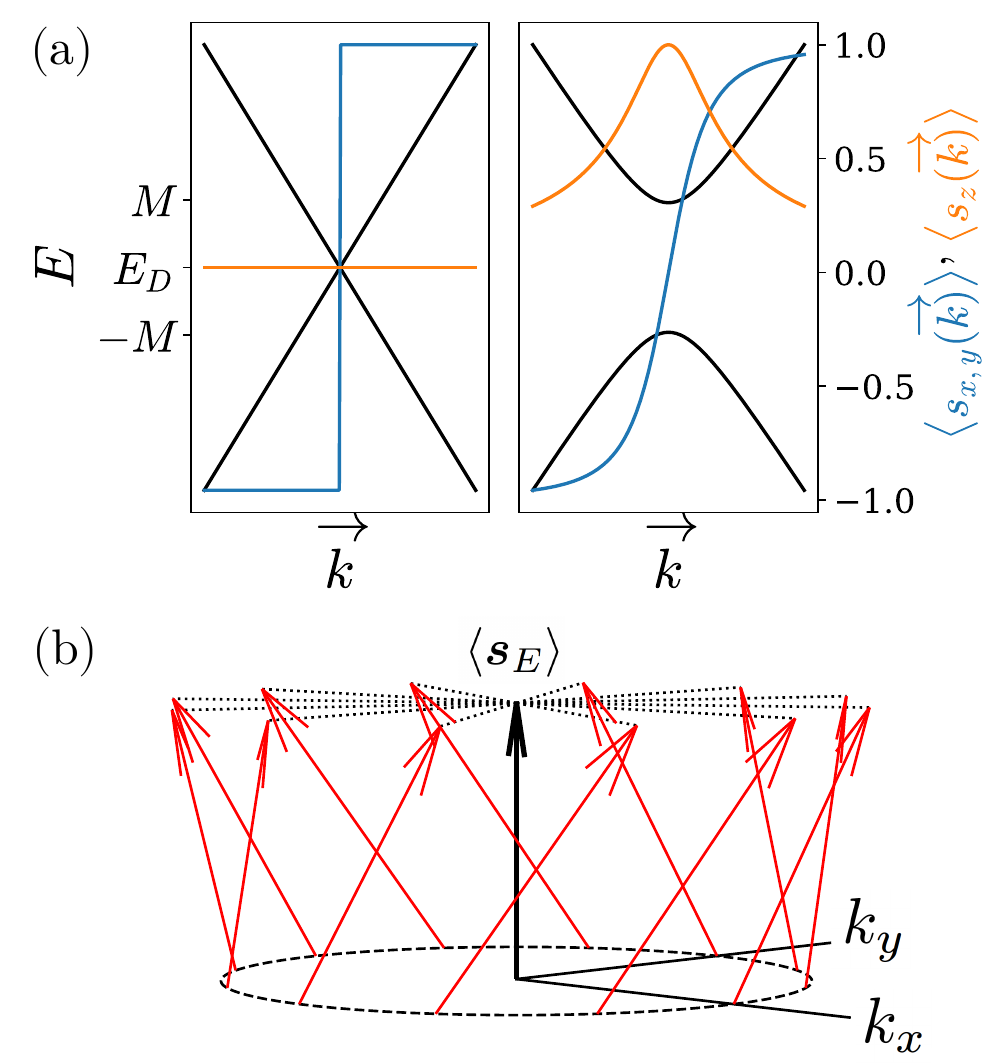}
\caption{(a) Band structure and spin textures (in units of $\hbar/2$) for the TI (left) and MTI (right) surface states. Only the conduction band spin textures are shown. The TI has a Dirac cone at energy $E_D$ while the MTI has gapped surface states with a gap of $2M$. (b) Spin texture (red arrows) of the MTI surface states for several k-points around the Fermi surface, depicted by a circular, dashed line. The average value of the spin texture (see Eq.\ \eqref{eq_SvsE}) is shown as a black line and only has $z$ component.}
\label{fig_F1}
\end{figure}

The spin dynamics and transport in the surface states of a magnetic topological insulator can be qualitatively understood with a two-band model describing a Dirac cone with exchange interaction \cite{PhysRevLett.102.156603,PhysRevB.82.161414} 

\begin{align}\label{eq_2band}
\mathcal{H} = \hbar v (\bm{k} \times \bm{s} )\cdot {\hat{\bm{z}}} + M s_z.
\end{align}
Here, $v$ is the Fermi velocity of the surface states, $\bm{k} = \left( k\cos(\theta),k\sin(\theta),0 \right)$ is the momentum with $k = |\bm{k}|$ and $\theta$ the polar angle, $\bm{s}=(s_x,s_y,s_z)$ is a vector of Pauli matrices acting on the spin and $M$ is the effective exchange interaction. When $M=0$, the band structure is that of a gapless Dirac cone typical of TI surface states, while a finite exchange interaction $M$ opens a gap in the spectrum and models the Chern insulator phase in MTIs. Such band structures are shown in Fig.\ \ref{fig_F1}(a).

Ubiquitous to spin transport in a given material is the spin texture of its Bloch states, which acts as an effective magnetic field and is usually band and momentum dependent,
\begin{align}
   \langle \bm{s}^n(\bm{k}) \rangle = \langle n\bm{k} | \bm{s} | n\bm{k} \rangle,
\end{align}
with $n$ being the band index. For Eq.\ \eqref{eq_2band}, the spin texture for the conduction and valence bands is
\begin{gather}\label{eq_2bandtexture}
    \langle \bm{s}^\text{CB}(\bm{k}) \rangle = -\langle \bm{s}^\text{VB}(\bm{k}) \rangle \\
    = \frac{1}{\sqrt{(\hbar v k)^2 + M^2}}
    \times (-\hbar v k \sin(\theta), \hbar v k \cos(\theta), M). \nonumber
\end{gather}
For $M=0$, the spin texture is that of a Rashba system with helical spin-momentum locking, while a nonzero $M$ induces a finite $\langle s_z(\bm{k})\rangle$. Importantly, the in-plane and out-of-plane components of the spin texture have opposite dependence on the magnitude of momentum, and therefore on energy. At low energies relative to the band edge, the spins are polarized almost completely along the $z$ direction, perpendicular to the TI surface. At high energies, $\langle s_z(\bm{k})\rangle$ decreases and the Rashba-like spin texture is mostly restored. This behavior can be observed in Fig.\ \ref{fig_F1}(a) (right panel), where we plot the spin texture as a function of momentum for the conduction bands. In Fig.\ \ref{fig_F1}(b) we plot the spin texture around the Fermi surface at a given energy $E$, highlighting the helical texture of the in-plane spins and the constant out-of-plane component.

As indicated by Eq.\ \eqref{eq_2bandtexture} and Fig.\ \ref{fig_F1}(b), the in-plane texture changes when scattering throughout the Fermi surface, while $\langle s_z (\bm{k}) \rangle$ remains constant. This has profound consequences for spin relaxation and transport in the MTI system, as energy-conserving scattering leading to a change in momentum is intimately related to spin relaxation due to the randomization of the $\bm{k}$-dependent spin texture \cite{Zutic2004}. In the framework of spin relaxation based on randomization of the effective magnetic field $\langle \bm{s}^n(\bm{k}) \rangle$, the spin will decay from a non-equilibrium distribution to its equilibrium state \cite{Kittel2004, Fabian2007}, given by the average value of the spin texture at the Fermi level,
\begin{equation} \label{eq_SvsE}
    \langle \bm{s}_{E} \rangle = \int_{FS} \langle \bm{s}^n(\bm{k}) \rangle \text{d} \bm{k}_F = (0,0,\frac{M}{E}).
\end{equation}

Although other spin relaxation mechanisms could be present in a system with magnetization (e.g.\ magnetic fluctuations), the above result suggests that out-of-plane spins should decay to a finite value and thus live much longer than in-plane spins, and this behavior should have a clear energy dependence. Therefore, measuring a large spin lifetime anisotropy of the gapped MTI surface states (not the chiral edge modes of the QAHE) in spin transport experiments could be used as a fingerprint to detect the Chern insulator phase of MTIs. This highlights the fact that in transport experiments, the topologically protected chiral edge states (i.e. the QAHE modes) are not the only ones carrying information about the topological phase of the MTI system, but the often ignored magnetization-induced gapped surface states can also be used to probe the Chern insulating nature of the material.

To corroborate our analysis, which is based on a simple two-band model, we now turn to numerical simulations of spin dynamics in a three-dimensional model of topological insulators with disorder. For that, we implement the Fu-Kane-Mele model in the diamond lattice with one orbital and spin per site, defined as \cite{Fu2007, Soriano2012,PhysRevLett.126.167701}
\begin{align}\label{eq_FKM}
    \mathcal{H}_\x{FKM} &= \sum_{\langle i,j \rangle, s} c_{i,s}^\dagger t_{ij} c_{j,s} \nonumber \\
    &+ \lambda \frac{8i}{a^2} \sum_{\langle \langle i,j \rangle \rangle, s, s'}  c_{i,s}^\dagger [\bm{s} \cdot (\bm{d}_{ik} \times \bm{d}_{kj})]_{s,s'} c_{j,s'} \nonumber \\
    &+ \sum_{i,s,s'} c_{i,s}^\dagger [\bm{M} \cdot \bm{s}]_{s,s'} c_{i,s'}.
\end{align}
The first term is a nearest-neighbor hopping connecting site $i$ in sublattice A with site $j$ in sublattice B. We define our unit cell containing sublattices A and B in positions $\bm{r}_\x{A} = (0,0,0)$ and $\bm{r}_\x{B} = a (0,0,\sqrt{3/2}/2)$, respectively, with $a$ the lattice constant. To obtain a topological insulator phase, $t_{ij}$ must not be equal for all neighbours \cite{Fu2007}; we choose $t_{ij}=t=1$ when the hopping connects site $i$ at $\bm{r}_\x{A}$ with sites $j$ at $\bm{r}_j = \bm{r}_\x{B} - \bm{a}_{1,2,3}$, and $t_{ij}=1.4t$ when the hopping couples sites $\bm{r}_\x{A}$ and $\bm{r}_\x{B}$. Here, $\bm{a}_{1,2,3}$ are the basis vectors $\bm{a}_1=a(1/2,-\sqrt{1/3}/2, \sqrt{2/3})$, $\bm{a}_2=a(0,\sqrt{1/3}, \sqrt{2/3})$, and $\bm{a}_3=a(-1/2,-\sqrt{1/3}/2, \sqrt{2/3})$. The second term is the next-nearest-neighbor spin-orbit coupling with strength $\lambda$, with $\bm{d}_{ik}$ and $\bm{d}_{kj}$ the vectors connecting sites $i$ and $j$ with their common neighboring site $k$. The third term is the effective exchange interaction akin to the second term in Eq.\ \eqref{eq_2band}. 

\begin{figure}[t] 
\centering
\includegraphics[width=0.48\textwidth]{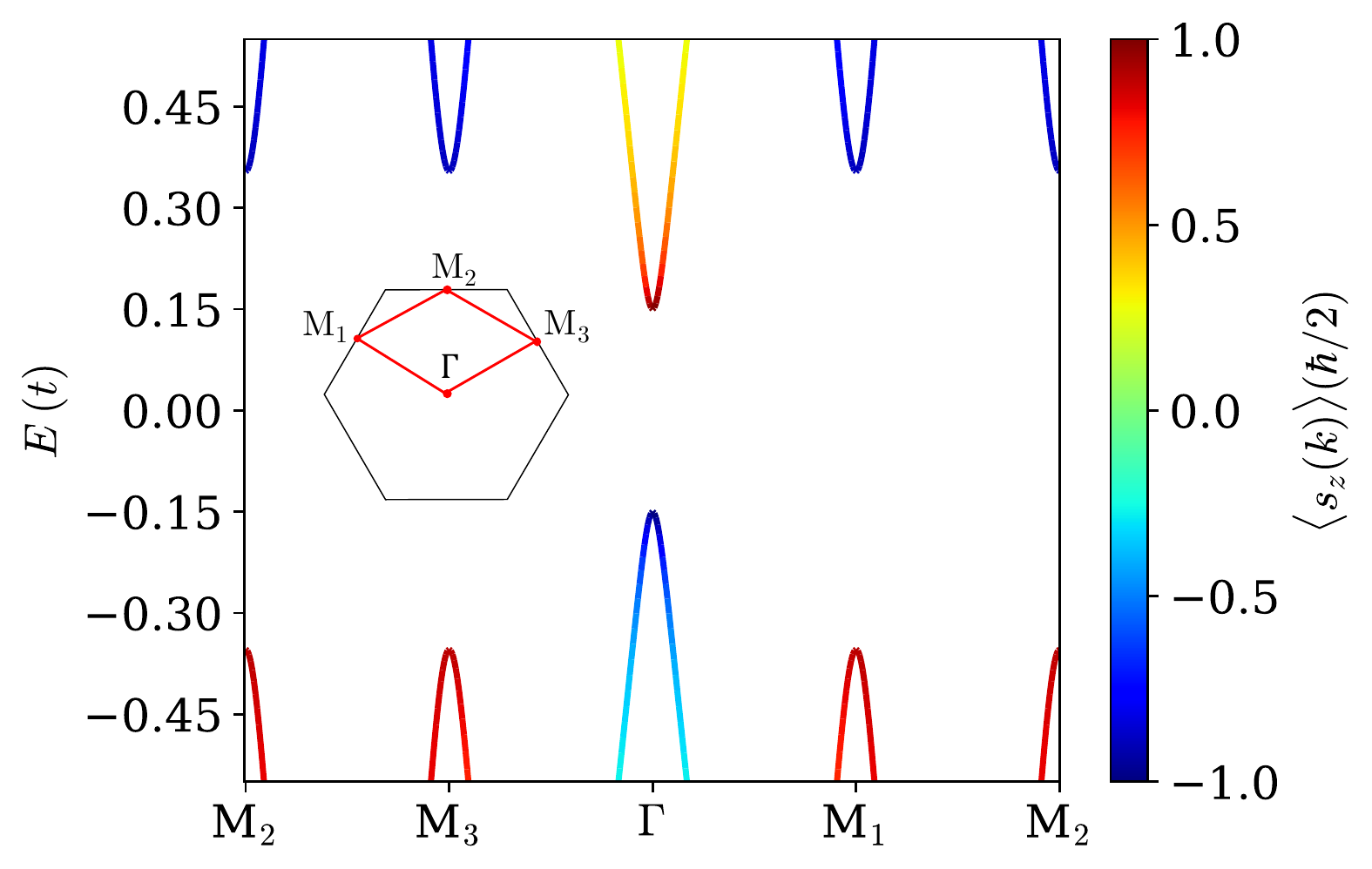}
\caption{Band structure of the tight-binding MTI model, Eq.\ \eqref{eq_FKM}, in a two-dimensional slab geometry with a thickness of 15 A-B dimer pairs. A gap of $2M = 0.3t$ in the top and bottom surface states is present due to the magnetization $M$. The color shows the $z$ component of the spin texture. Inset: Brillouin zone and high-symmetry points of the two-dimensional slab hexagonal lattice.}
\label{fig_F2}
\end{figure}

This model was shown to generate the quantum anomalous Hall insulator phase exhibiting chiral edge states when realized in a two-dimensional slab geometry \cite{PhysRevLett.126.167701}. Thus, we also implement Eq.\ \eqref{eq_FKM} in a slab geometry with a finite thickness of 15 A-B site pairs along the $z$ direction and periodic boundary conditions in the $xy$ plane. This thickness is enough to prevent spurious gap opening due to hybridization of the top and bottom surfaces \cite{Soriano2012}. Without loss of generality, we choose $\lambda = 0.125t$ and $\bm{M} = (0,0,M)$, with $M=0.15t$, noting that the exchange interaction is perpendicular to the surface so a gap in the topological surface states can be opened, realizing the Chern insulator. We present the resulting two-dimensional band structure in Fig.\ \ref{fig_F2}. The states at low energy ($|E|<0.35t$) correspond to the gapped surface states of MTIs, being doubly degenerate due to the existence of both top and bottom surfaces. As expected, a gap of $2M$ is present, and the conduction and valence bands display a spin texture very similar to that of Eq.\ \eqref{eq_2bandtexture}: near the band edge, spins are completely polarized along the $z$ direction and this value decreases with increasing energy. We note that at energy $E \approx 0.35 t$ the bulk states (i.e.\ not from the surfaces) appear, and that the chiral states from the QAHE are not present because of the periodicity in the $x$ and $y$ directions.

Next, to study spin relaxation, we proceed in calculating the time evolution of spin-polarized states projected onto energies corresponding to the MTI gapped surface states. The time- and energy-dependent spin expectation value is given as \cite{Tuan2014,Fan2021}
\begin{align}\label{eq_spinLSQT}
    \langle \bm{S}(E,t) \rangle = \Re \frac{\langle \Psi(t) |\bm{s} \delta(E-\mathcal{H}_\x{FKM}) \Psi(t)\rangle}{\langle \Psi(t) |\delta(E-\mathcal{H}_\x{FKM})| \Psi(t)\rangle}.
\end{align}
The computation of the projector operator $\delta(E-\mathcal{H}_\x{FKM})$ is performed with the kernel polynomial method, as detailed in Ref.\ \citenum{Fan2021}, and spin-polarized random-phase vectors are employed to prepare the initial state $|\Psi(0)\rangle$ which is then evolved as $|\Psi(t)\rangle = e^{-i \mathcal{H}_\x{FKM} t / \hbar} | \Psi (0) \rangle$. The evaluation of $\delta(E-\mathcal{H}_\x{FKM})$ is carried out in Kwant \cite{Groth2014} upon the real-space implementation of Eq.\ \eqref{eq_FKM} with Pybinding \cite{dean_moldovan_2020_4010216}. A supercell of size $75a \times 75a$ is used in the $xy$ plane, 500 moments are employed in the expansion of $\delta(E-\mathcal{H}_\x{FKM})$, and 5 random vectors are used. Finally, to include scattering in the simulation and describe a more realistic transport regime, to Eq.\ \eqref{eq_FKM} we add random Anderson disorder $\sum_i c_i^\dagger U_i c_i$ with $U_i$ uniformly distributed in the interval $[-U/2,U/2]$.

\begin{figure}[tb] 
\centering
\includegraphics[width=0.48\textwidth]{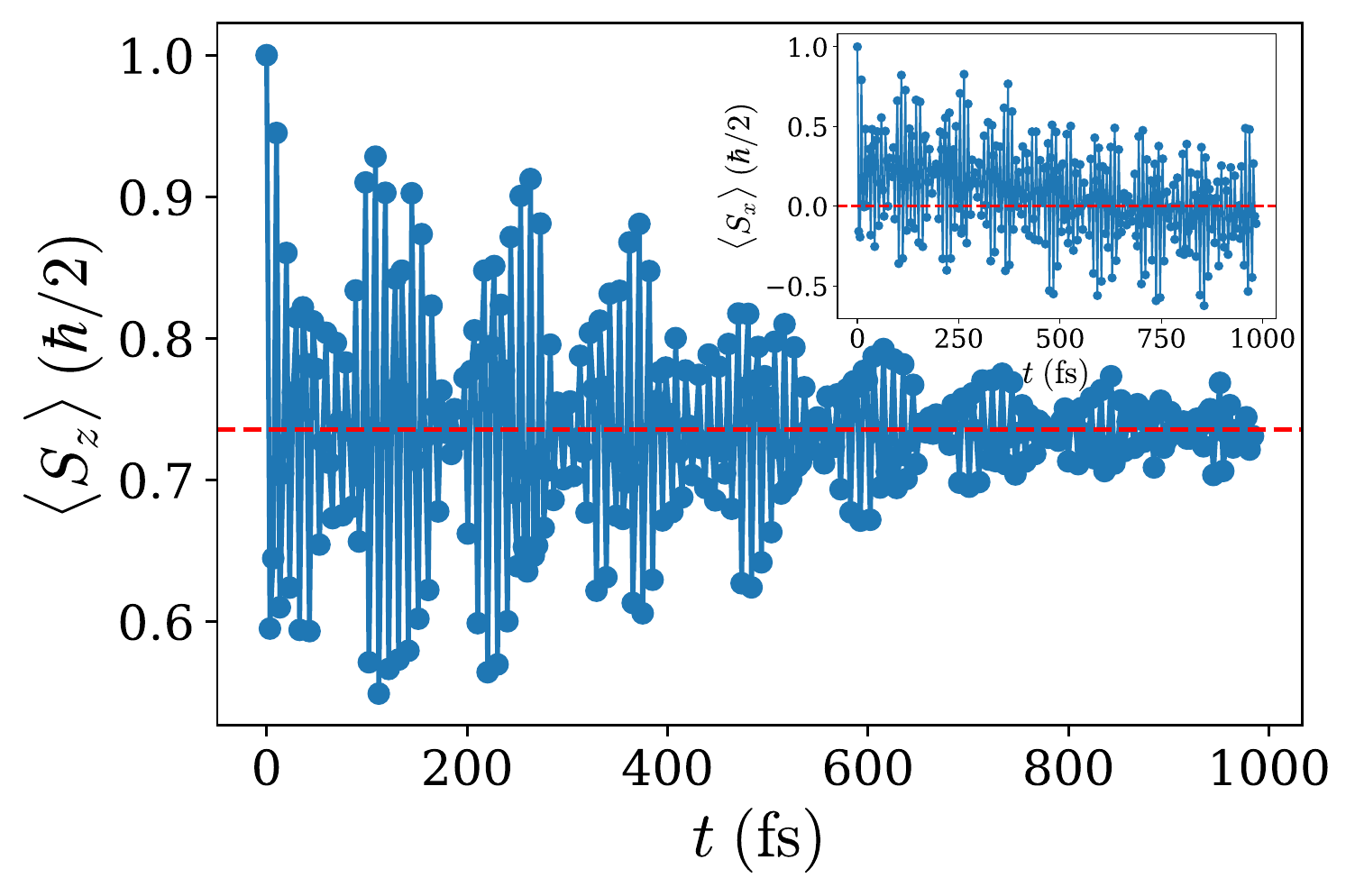}
\caption{Time evolution of the spin expectation value (Eq.\ \eqref{eq_spinLSQT}), for a state initially polarized along $z$. The red dashed line highlights the remanent spin polarization that does not relax. Inset: same as the main plot, but for a state initially polarized along $x$. In both cases, $E = 0.2t$ and $U = 0.2t$.}
\label{fig_F3}
\end{figure}

We plot in Fig.\ \ref{fig_F3} the time evolution of the out-of-plane spin of a state initially polarized along the $z$ direction. We consider $E = 0.2t$ (i.e., close to the band edge) with a disorder strength $U = 0.2t$ that is smaller than the topological gap.
We see three features that are characteristic of spin dynamics in this MTI system. First, the spin precesses around the effective magnetic field with angular frequency $\omega = 2|E|/\hbar$ \cite{Liu2013, Cummings2016}, resulting in the fast oscillation of $\langle s_z \rangle$. Second, the oscillating part of the spin quickly decays over a time scale of $\sim$$1$ ps, indicating very fast dephasing arising from charge scattering \cite{Liu2013} and energy broadening \cite{Cummings2016}. However, the most remarkable part is that even in the presence of disorder, the spin decays to a finite remanent value that does not appear to relax, even when the oscillating component has practically vanished. The red dashed line indicates this finite remanent spin, which coincides well with the value $M/E = 0.75$ predicted by Eq.\ \eqref{eq_SvsE}. Thus, even in the presence of disorder, the uniform out-of-plane spin texture of the MTI surface states appears to lead to very long spin lifetimes that may be detectable experimentally.

In the inset of Fig.\ \ref{fig_F3} we plot the same for in-plane spins, with an initial polarization along $x$. Here we see similar features, except that $\langle s_x \rangle$ decays to zero. This behavior is also consistent with the prediction of Eq.\ \eqref{eq_SvsE}, as charge scattering will randomize momentum over the entire Fermi surface, which, with the helical in-plane spin texture leads to zero average in-plane spin. Thus, in the MTI system, we observe very short in-plane spin lifetimes, while out-of-plane spins decay to a remanent value that is stable over very long times.

\begin{figure}[tb] 
\centering
\includegraphics[width=0.48\textwidth]{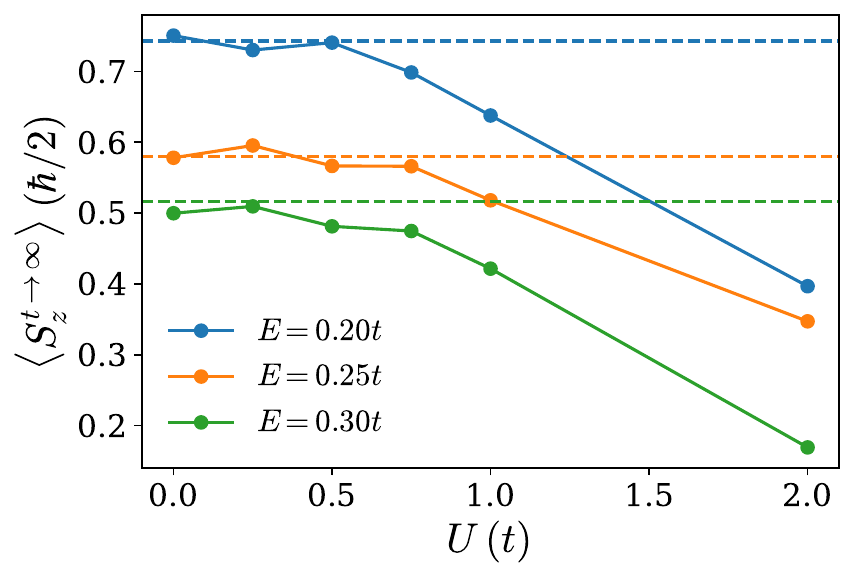}
\caption{Remanent spin polarization as a function of disorder strength at different Fermi levels. Dashed lines are the predicted values derived from the two-band model (Eq.\ \eqref{eq_SvsE}).}
\label{fig_F4}
\end{figure}

In Fig.\ \ref{fig_F4}, we plot the magnitude of the remanent out-of-plane spin polarization as a function of disorder strength for different energies. There are two important trends to note here. First, the remnant of $\langle S_z^{t\rightarrow\infty} \rangle$ decreases with increasing energy, in accordance with the intuition gained from the two-band model. Indeed, the dashed lines in Fig.\ \ref{fig_F4} are the values of remanent spin polarization predicted from Eq.\ \eqref{eq_SvsE}, which match very well with our numerical simulations of the FKM model at low disorder when only the surface states are occupied. Second, although the robust out-of-plane spin signal decreases with increasing disorder, it remains finite even for very strong disorder nearly $10\times$ larger than the magnetic gap. This strong resilience to disorder suggests that measurements of long out-of-plane spin lifetime (and large spin lifetime anisotropy) may serve as an experimental fingerprint of the MTI phase.

We have identified unprecedented spin transport features in MTIs, which are resilient to electrostatic disorder and which could serve as an experimental smoking gun of the formation of the Chern insulator phase. Experiments based on Hanle spin precession measurements, as implemented in other systems such as graphene proximitized with magnetic materials \cite{Karpiak2020, Ghiasi2021}, should reveal such a long lasting and energy-dependent out-of-plane spin lifetime as well as a large spin lifetime anisotropy \cite{Raes2016}. In MTIs, the attention is often focused on the nontrivial chiral edge states. Our results show that the Chern insulator phase could be also probed through the overlooked gapped surface states. MTIs are usually obtained either via magnetic doping of the TI bulk or by magnetizing only the surfaces via proximity effect. Our simulations found identical results in both cases (by adding the exchange interaction only in the two outermost dimer pairs, or throughout the entire TI structure), making the reported spin transport features robust to the experimental design. Additionally, we also obtained identical results whether injecting spin-polarized states into one or both layers. Therefore, one could envision injecting spins via either a side or a top magnetic contact \cite{Xu2018, Karpiak2018}. While the former would incorporate spins into both surface states, the latter would only couple to a single surface, and both approaches could be used to probe the spin lifetime anisotropy in the MTI phase. In the future one could also envision studying spin dynamics phenomena in MTIs in a broader context, including the search for more exotic quantum effects, as topological materials have been proposed to provide a scaffold for exploring axion physics and dark matter by Wilczek \cite{PhysRevLett.58.1799}, fostering the field of axion electrodynamics in condensed matter \cite{10.1063/5.0038804,doi:10.1126/sciadv.aao1669,Gao2021,LI20231252,PhysRevB.106.195157}.

\begin{acknowledgments}
\textit{Acknowledgements ---} We thank Pablo M.\ Perez-Piskunow, Imen Taktak and Ewelina Hankiewicz for fruitful discussions. M.V.\ was supported as part of the Center for Novel Pathways to Quantum Coherence in Materials, an Energy Frontier Research Center funded by the US Department of Energy, Office of Science, Basic Energy Sciences. We acknowledge the European Union Horizon 2020 research and innovation programme under Grant Agreement No. 824140 (TOCHA, H2020-FETPROACT-01-2018). ICN2 is funded by the CERCA programme / Generalitat de Catalunya, and is supported by the Severo Ochoa Centres of Excellence programme, Grant CEX2021-001214-S, funded by MCIN/AEI/10.13039.501100011033.
\end{acknowledgments}


%

\end{document}